\documentclass[%
superscriptaddress,
 amsmath,amssymb,
 aps,
longbibliography
]{revtex4-2}
\usepackage[utf8]{inputenc}
\usepackage{graphicx}
\usepackage{epstopdf}
\usepackage{amsmath,amssymb}
\usepackage{xcolor}
\usepackage{graphicx}
\usepackage{dcolumn}
\usepackage{bm}
\usepackage{hyperref}

\begin{document}

\preprint{APS/123-QED}

\title{
Imprints of log-periodicity in thermoacoustic systems close to lean blowout}
\author{Ankan Banerjee}
\email{ankan1090@gmail.com}
\affiliation{Department of Aerospace Engineering, Indian Institute of Technology Madras, Chennai 600036, India}  
\author{Induja Pavithran}%
\affiliation{Department of Aerospace Engineering, Indian Institute of Technology Madras, Chennai 600036, India}
\affiliation{Department of Physics, Indian Institute of Technology Madras, Chennai 600036, India}
\author{R. I. Sujith}%
\affiliation{Department of Aerospace Engineering, Indian Institute of Technology Madras, Chennai 600036, India}%

\date{\today}
\begin{abstract}
In the context of statistical physics, critical phenomena are accompanied by power laws having a singularity at the critical point where a sudden change in the state of the system occurs. In this work, we show that lean blowout (LBO) in a turbulent thermoacoustic system can be viewed as a critical phenomenon. As a crucial discovery of the system dynamics approaching LBO, we unravel the existence of the discrete scale invariance (DSI). In this context, we identify the presence of log-periodic oscillations in the temporal evolution of the amplitude of dominant mode of low-frequency oscillations $(A_f)$ exist in pressure fluctuations preceding LBO. The presence of DSI indicates the recursive development of blowout. Additionally, we find that $A_f$ shows a faster than exponential growth and becomes singular when blowout occurs. We then present a model that depicts the evolution of $A_f$ based on log-periodic corrections to the power law associated with its growth. Using the model, we find that blowout can be predicted even several seconds earlier. The predicted time of LBO in good agreement with the actual time of occurrence of LBO obtained from the experiment.
\end{abstract}
\maketitle

\section{Introduction}
Lean premixed combustion is one of the most sought-after technologies in gas turbine engines to satisfy the stringent emission norms of oxides of nitrogen (NOx)~\cite{turns_pollutant_book:2000}. The lean fuel-air mixture results in a significant reduction of the flame temperature inside the combustor near the reactant flame due to the presence of excess air. However, such lean conditions make an engine susceptible to lean-flame blowout. Lean-flame blowout (LBO), continues to be detrimental to the operation of modern gas turbine combustors in the power-production and aviation industries~\cite{lieuwen2021unsteady}. Blowout is the state when the flame fails to stabilize inside the combustor and gets blown out due to the reduced flame speed in comparison to the high flow rates of incoming  reactants~\cite{plee_CAF:1979,radhakrishnan_CAF:1981}. In power plants based on gas turbine engines, a blowout can lead to unplanned power outages and increased operational costs~\cite{nair_JPP:2005}. Both military and commercial aircraft engines are also susceptible to blowout during lean operation and sudden changes in throttle settings~\cite{s_nair_thesis:2006}. Therefore, precursors to blowout are desired to avoid unplanned shut down of engines in aircrafts and power plants.

In order to understand the physical mechanism behind the occurrence of a blowout, a number of studies have analyzed the flame dynamics prior to its occurrence for different combustors and different flame holding mechanisms~\cite{nair_JPP:2005, nair_JPP:2007,chaudhuri_CAF:2008,chaudhuri_combustion_flame:2010, muruganandam_IJSCD:2012,unni_chaos:2018}. Nair and Lieuwen~\cite{nair_JPP:2005, nair_JPP:2007} analyzed flame dynamics in a turbulent premixed combustor near blowout. They characterized blowout as a consequence of two phenomena; the emergence of localized flame extinction regions or flame holes followed by the violent flapping of the flame front or collective extinction of flames. The amplitude of acoustic oscillations increases due to such flame behavior. As the system approaches blowout, they observed bursts in the amplitude of the acoustic pressure field. These bursts correspond to the extinction and reignition of flames. Later, Chaudhury et al.~\cite{chaudhuri_combustion_flame:2010} investigated the interaction between flame fronts and shear layer vortices prior to blowout in a turbulent premixed combustor with a bluff body stabilized flame and showed that the interaction leads to local extinction of flames or generation of local (smaller) flame holes. The number of flame holes and the frequency of their appearance increase as the system approaches blowout. The collective behavior of smaller flame holes leads to the formation of a global flame hole that causes blowout. The formation of flame holes that are localized in space can be considered as a small-scale intermediate state to blowout. Such intermediate states have been used to control blowout and produce warning signals~\cite{muruganandam_active_control_LBO:2005}. Several other phenomena, such as earthquakes and stock market crashes, are found to be preceded by similar small-scale precursory events. Studies have shown that large earthquakes~\cite{sornette_PAG:1994} or stock market crashes~\cite{Johansen_2000} are forms of self-organized criticality, and they occur as a scaling-up process of its earlier precursory events. The hierarchical dynamics underlying the precursory events were utilized to predict the corresponding critical phenomena based on the property of scale-invariance~\cite{sornette_phys_report:1998}.

The property of scale-invariance means that the law governing a physical variable of a system remains invariant under the change of a scale (in length, time, energy or some other variables) by some common factor. In the case of critical phase transition, the scale-invariance property exists in the close vicinity of a critical point. A physical variable of the system showing such a property follows a power law with a real exponent in a local neighborhood (the so-called asymptotic critical region) of the critical point~\cite{saleur_JGRSE:1996}. The variable is said to have continuous scale-invariance property if the exponent of the power law is real. Moreover, as the system approaches the critical point, the variable becomes singular at the critical point. In other words, the scale-invariance property ceases to exist beyond the critical point following a singularity there~\cite{saleur_JGRSE:1996}. 

Although the continuous scale-invariance property can explain a transition occurring at the critical point, it has a limited scope in detecting precursory signals outside the asymptotic critical region. This limitation has been overcome by extending the exponent in the power law from a real-valued to a complex-valued exponent. Complex-valued exponents are associated with discrete scale-invariance (DSI) and result in log-periodic oscillations to the scaling law (detailed discussions are given in Section~\ref{subsec:log-periodicity}). A power law decorated with the log-periodic correction is called a \textit{log-periodic power law} (LPPL)~\cite{sornette2009_book}. LPPL has successfully predicted critical transitions in several natural complex systems such as earthquakes~\cite{Sornette_1995}, icequakes~\cite{faillettaz_icequake:2008} as well as in complex human-made systems such as rupture stresses from acoustic emissions~\cite{Anifrani_1995}, stock market crashes~\cite{Sornette_1996, Johansen_2000}, credit risk estimation~\cite{wosnitza_EPJB:2015} to name a few. The strength of the LPPL formulation is that it can predict an impending critical transition by detecting its implicit precursory phenomena, which are not identified by pure power law formulation associated with continuous scale-invariance in the critical region.
 
Detection of precursory signals to an impending blowout for providing efficient early warnings is highly desirable so that a combustor can approach leaner conditions without risking blowout. Thermoacoustic systems possess inherent complexity due to the nonlinear interaction between their subsystems, namely, the acoustic field, the turbulent hydrodynamic field and the flame~\cite{sturgess_EGTP:1997,ahmed_CST:2017,wang_AST:2019}. The complexity of the system intrigues researchers to investigate its dynamical behaviors, which, in turn, can help identify precursors to an impending blowout. Subjects such as nonlinear dynamics~\cite{strogatz_nonlinear_book:2018}, complex systems theory~\cite{bar1998dynamics} and pattern formation~\cite{hoyle_patern_book:2006} have been utilized significantly in this regard.

Gotoda and his collaborators used tools from nonlinear dynamics to analyze the system behavior prior to lean blowout~\cite{gotoda_chaos:2011, gotoda_CINS:2012, gotoda_PRE:2014}. The signature of self-affine structures~\cite{gotoda_CINS:2012} in the dynamics near lean blowout (LBO), and the translational error~\cite{gotoda_PRE:2014}, were used as precursors to detect LBO. Mukhopadhyay et al.~\cite{Mukhopadhyay_JPP:2013} proposed a precursor for LBO based on symbolic time series analysis. Multifractal characteristics of pressure fluctuations quantified by the Hurst exponent have been used as early warning signals to impending thermoacoustic instability and LBO~\cite{nair_JFM:2014, unni_JFM:2015}. Recurrence quantification analysis (RQA) is another technique providing precursor measures that show distinctive signatures toward LBO~\cite{unni_RQA:2016, de_chaos:2020}. Recently, Bhattacharya et al.~\cite{bhattacharya_ATE:2020} proposed a fast Fourier transform (FFT) based single scalar-valued measure to detect different operational regimes, namely, stable operation, thermoacoustic instability (TAI), and LBO based on the time series of acoustic pressure. Detailed discussions about the mechanism of LBO, its precursors, and control are summed up well in review articles~\cite{shanbhoguePECS:2009, sujith_PCI:2021}.

In earlier experiments~\cite{muruganandam_active_control_LBO:2005,nair_JPP:2005,nair_JPP:2007,chaudhuri_CAF:2008,chaudhuri_combustion_flame:2010,unni_JFM:2015,unni_chaos:2018}, warnings to an impending blowout and its control were dependent on user defined threshold values of an underlying property of the associated system. Such kind of threshold values are system dependent and act as a constraint in reaching leaner fuel-air ratio. Moreover, the control parameter, the fuel-air ratio, was changed in a quasi-static manner and the system was let to stabilize for each control parameter. In other words, those systems were treated as autonomous. However, real-life combustion systems are mostly non-autonomous. The control parameter changes constantly at a finite rate which makes the analysis of the system much more challenging than an autonomous one. 
Therefore, in the case of a non-autonomous thermoacoustic system, it would be fascinating to investigate whether there are potential precursory signals for predicting LBO. Additionally, rather than any threshold-dependent methods, prediction of the time to LBO will be more convenient in non-autonomous systems for circumventing LBO. In the present work, we attempt to address these issues by interpreting LBO as a critical phenomenon from the perspective of critical phase transitions in statistical physics. We then use the LPPL formulation to predict the time to LBO several seconds in advance.

\section{\textbf{Log-periodicity in discrete scale-invariance}} \label{subsec:log-periodicity}
In Euclidean geometry, the dimension of a system or, the number of independent vectors (or bases) representing the system, is a positive integer. Mandelbrot generalized the concept of Euclidean geometry by introducing the concept of fractals which are \textit{sets consisting of parts similar to the whole}~\cite{mandelbrot_book:1982fractal}. Systems with fractals are said to have non-integer dimensions. Systems having non-integer dimensions possess the property of scale-invariance, which is represented by the equation
\begin{equation}
    F(x)=\mu F(\lambda x),
    \label{eq:scale_invariance_relation}
\end{equation}
where $x$ is a variable representing a scale of length, time, energy, etc. (mathematically $x$ generates a scale), $\lambda$ is a non-zero number known as the scale factor, and $\mu$ is a function of $\lambda$. Both $\lambda$ and $\mu$ are real numbers, and $F$ is a function associated with a physical variable of the system (or, $x$ is a measuring variable and $F(x)$ is a measured variable). A power law expressed as
\begin{equation}
    F(x)=Cx^\alpha,~\text{with}~\alpha=-\frac{\log~\mu}{\log~\lambda}
     \label{eq:power_law_continuous}
\end{equation}
is a solution to Eq.~(\ref{eq:scale_invariance_relation}). Note that $\alpha$ in this power law gives the fractal dimension. In the case of continuous-scale invariance, $\alpha$ is real. The scale invariance is called continuous since $\lambda$ can be any arbitrary real number. The scale-invariance property exists in an abundance of natural phenomena showing fractals, criticality, or self-organized criticality~\cite{bak_book:2013}. 

If the arbitrariness of scale factors is constrained in such a way that the scale factors are determined by a unique non-zero real number $\lambda$ and belong only to the set $ S=\{\lambda^i: i \text{ is an integer},~\lambda \neq 0\} $, then the scale-invariance property~(Eq.~(\ref{eq:scale_invariance_relation})) is said to be the discrete-scale invariance (DSI). In other words, the scale-invariance property holds if the scale factors are in geometric progression in $\lambda$ and appears periodically at each scale factor that belongs to $S$. Note that, the unique value of $\lambda$ is system dependent. The continuous change of the variable $x$ in the case of DSI gives the power law solution to Eq.~(\ref{eq:scale_invariance_relation}) of the form, 
\begin{equation}
    F(x)=Cx^\alpha P(\frac{\log x}{\log \lambda}).
     \label{eq:power_law_discrete}
\end{equation}
By expanding the periodic term $P(\frac{\log x}{\log \lambda})$ into a Fourier series, we get
\begin{eqnarray}
    F(x) &=& Cx^\alpha \sum A_n \exp{(i 2 n \pi \frac{\log x}{\log \lambda})}, \nonumber \\
      &=& C \sum A_n x^\alpha x^{i \frac{ 2 n \pi}{\log \lambda}}, \nonumber \\
      &=& C \sum A_n x^{\alpha +i n \omega},
     \label{eq:power_law_discrete_final}
\end{eqnarray}
where,  $\omega=2 \pi/\log\lambda$. Therefore, DSI leads to a complex fractal dimension $\alpha+i n \omega$. Moreover, it can be shown that the log-periodic oscillation stems from the imaginary part ($n \omega$) of complex fractal dimensions. For simplicity, we shall restrict ourselves to the first harmonic of the Fourier series. In that case, Eq.~(\ref{eq:power_law_discrete_final}) can be rewritten as,
\begin{eqnarray}
 F(x) &=& C \sum_{n=-1}^{n=1} A_n x^{\alpha+i\omega_n} \nonumber \\
 &=& C x^\alpha (A_0 + 2A_1 \cos(\omega \log x)),
\label{eq:power_law_log_periodicity}
\end{eqnarray}
where $A_1=A_{-1}$ and $\omega=2 \pi/\log\lambda$. Due to the periodic term in $\log x$, discrete-scale invariance is said to have a log-periodicity. Here, $\omega$ is the angular frequency, determined by $\lambda$ and has a unique value for a specific DSI. Since scale factors follow a geometric progression, DSI can represent hierarchical systems. DSI has been observed in theoretical systems such as Cantor sets, hierarchical diamond lattices, idealized Ising models etc~\cite{derrida_JSP:1983,meurice_PRL:1995}. Additionally, DSI has been found to exist in several heterogeneous and irreversible phenomena such as earthquakes, and stock market crashes~\cite{sornette_phys_report:1998}.

The presence of DSI in heterogeneous systems implies that critical phenomena can be viewed as hierarchical phenomena. For example, Sornette and Sammis~\cite{Sornette_1995} proposed that the occurrence of a large earthquake is a consequence of the propagation and accumulation of several preceding smaller earthquakes and ruptures in a large geographical area. A similar kind of behavior has also been shown to precede stock market crashes. Johansen et al.~\cite{Johansen_2000} hypothesized that stock market crashes are caused by the slow buildup of long-range correlations leading to the collapse of the stock market in one critical instant. Thus, DSI provides additional constraints on a system (in terms of a preferred scale factor) eventually unravelling the underlying physical mechanism.

\section{Experiments}
\begin{figure*}[ht]
    \centering
    \includegraphics[height=!,width=0.8\textwidth]{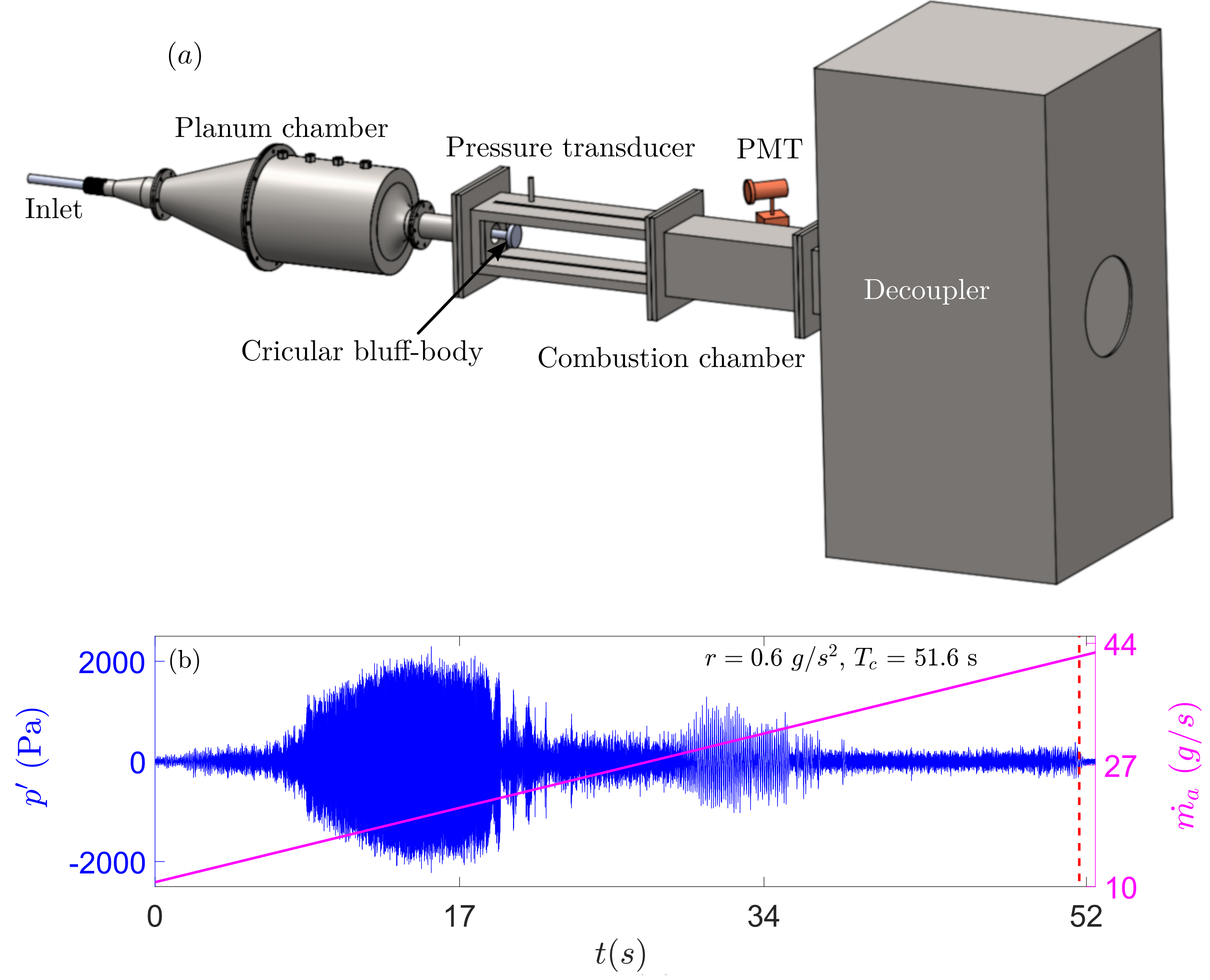}
    \caption{ (a) Schematic of the turbulent combustor used in the current study. Pressure fluctuations are measured using a piezoelectric transducer. The design of the combustor was adapted from Komarek and Polifke~\cite{komarek_combustor_design:2010}. (b) Temporal variation of the mass flow rate of air (magenta curve) together with the time series of pressure fluctuations (blue curve) for the rate of change of air flow rate $a=0.6~g/s^2$ are shown. The occurrence of blowout is indicated by the red dashed line.}
    \label{fig:schematic_diagram}
\end{figure*}

Experiments are performed on a turbulent combustor at high Reynolds numbers ($Re>14,000$) with a circular bluff-body as the flame-holding mechanism. A schematic diagram of the experimental setup is portrayed in Fig.~\ref{fig:schematic_diagram}(a). The system comprises a plenum or settling chamber and a combustion chamber with extension ducts. Fluctuations in the inlet air are diminished in the plenum chamber. A circular disc of radius 47 mm and thickness 10 mm is mounted on the central shaft as a bluff-body. The fuel, liquefied petroleum gas ($60~\%$ butane and $40~\%$ propane), is injected 100 mm upstream of the dump plane. The central shaft is used to deliver fuel into the combustor through four radial injection holes of diameter 1.7 mm in the central shaft.

The combustion chamber is cuboid in shape with a cross-section of size 90 mm $\times$ 90 mm and length 700 mm. A spark plug driven by a step-up transformer is mounted near the dump plane for ignition of the fuel–air mixture. Mass flow controllers (Alicat Scientific, MCR Series) are used to measure and control mass flow rates of air $(\Dot{m_a})$ and fuel $(\Dot{m_f})$ with an uncertainty of $\pm~(0.8~\%$ of the reading $+0.2~\%$ of the full scale). 
The Reynolds number $(Re)$ for the reactive flow is obtained as $Re = 4 \Dot{m} /\pi \nu (D_0+D_1),$ where $\Dot{m}=\Dot{m_a}+\Dot{m_f}$ is the mass flow rate of air-flow mixture, $(D_0)$ is the diameter of the burner, $(D_1)$ is the diameter of the circular bluff-body and $\mu$ is the dynamic viscosity of the air-fuel mixture in the experimental conditions. The Reynolds number for the reported experiments are varied within a range from $Re=(1.41 \pm 0.08) \times  10^4$ to $Re=(5.3  \pm 0.26) \times 10^4$. We measure the acoustic pressure fluctuations in terms of voltage ($V$) using a piezoelectric sensor (PCB103B02, sensitivity: 217.5 mV/kPa, resolution: 0.2 Pa and uncertainty: 0.15 Pa) at a sampling rate of 12 kHz. The global heat release rate is measured from the CH* chemiluminescence intensity~\cite{hardalupas_CAF:2004}, which is captured using a photomultiplier tube  (PMT) (Hamamatsu H10722-01) outfitted with a bandpass filter (wavelength 435 nm and 10 nm full width at half maximum). More details of the experimental setup are discussed in~\cite{raghunathan_JFM:2020}.

In the present study, the global equivalence ratio $\phi,~(\phi=\frac{({\Dot{m_f}/\Dot{m_a})}_{actual}}{({\Dot{m_f}/\Dot{m_a})}_{stoichiometry}})$
is decreased as we increase the air flow rate while keeping the fuel flow rate constant, $1.07~g/s$. The air flow rate (the control parameter) is varied linearly with respect to time ($t$) at a constant rate $r$. The equivalence ratio is varied from 1 to 0.25 continuously in time in each experiment. We perform experiments for different values of $r$ ranging from $0.1~\text{to}~2.0 ~g/s^2$. variation of acoustic pressure in time, as we vary the airflow rate is shown in Fig.~\ref{fig:schematic_diagram}(b) for $r=0.6 ~g/s^2$. The blue curve in the plot represents acoustic pressure fluctuations $(p^\prime)$ (in Pa) while the change of airflow rates is exhibited by the magenta curve. Initially, the thermoacoustic system is in a state of stable operation, and the amplitude of $p^\prime$ is close to $100$ Pa. The amplitude of pressure fluctuations $(p^\prime)$ increases with the airflow rate ramping up at a constant rate $r$. The system experiences thermoacoustic instability due to positive feedback between the acoustic oscillations and the unsteady heat release rate~\cite{sujith_ARFM:2018}. The amplitude of $p^\prime$ reaches its maximum value during thermoacoustic instability. Further increment in airflow rates (i.e. reducing the equivalence ratio) results in the reduction of the amplitude of pressure fluctuations, and the combustor approaches blowout. The occurrence of blowout is determined as the instance at which heat release rate becomes zero, and is represented by the red dashed line ($T_c$) in Fig.~\ref{fig:schematic_diagram}(b). In the rest of the paper, we focus on data close to blowout to serve our purposes.

\section{Results}
We observe from experiments the presence of low-frequency oscillations ($\approx 10~Hz$) among other high-frequency fluctuations and aperiodicity, before the system goes to the blowout state. The appearance of low-frequency oscillations prior to blowout has been reported in the literature~\cite{nair_JPP:2005,yi_AIAAJ:2007}, and have been used to characterize blowout where control parameters were changed in a quasistatic manner. Here, in experiments with continuously varying parameter, we also use those low-frequency oscillations for characterizing blowout. As a result, we decompose $p^\prime$ into Fourier modes and construct a new time series from Fourier coefficients. Since we are varying the control parameter continuously, we perform the Fourier transform of $p^\prime$ window wise. We consider one second windows with an overlap of $0.9~s$ for determining the amplitude spectrum at each time instance ($t_i~(s)$). During this process, we identify a significant presence of a low-frequency spectrum with frequencies ($f$) from $5$ to $25~Hz$ over the entire parameter range ($0.1<r<2.0~g/s^2$) explored in the present article. Therefore, in order to construct the desired time series, we define a new variable $A_f(t_i)=\max(C_f(f,t_i))$, where $C_f(f,t_i)$ are Fourier coefficients of $f$ for each $t_i$ computed over the time window $[t_i-1,t_i]$.

\begin{figure}[h!]
    \centering
    \includegraphics[height=!,width=0.45\textwidth]{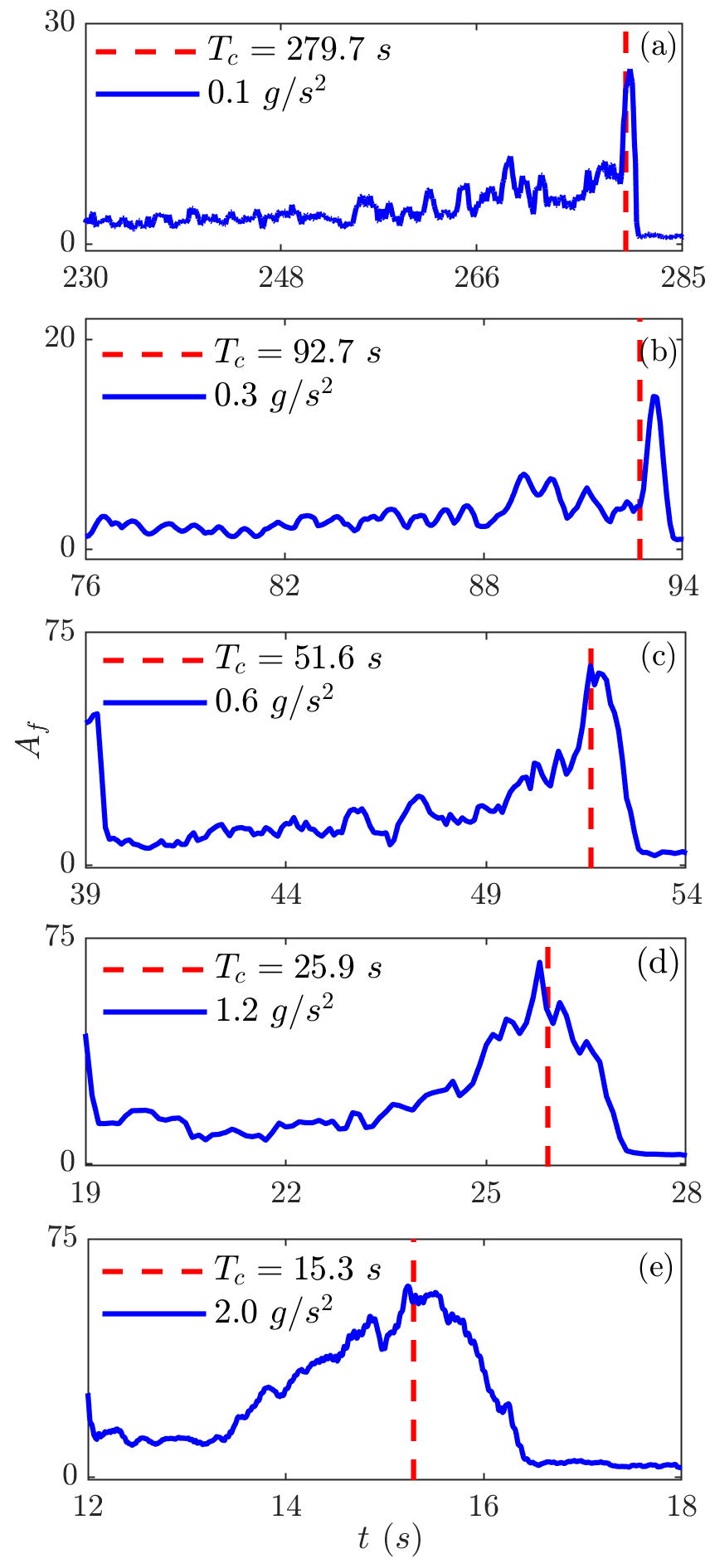}
    \caption{Time series of the maximum of the amplitude spectrum ($A_f$) of low-frequency ($5<f<25$) as obtained from experiments corresponding to different rates of change of airflow rates. Red dashed lines represent critical time ($T_c$) for blowout obtained from heat release rate.
    }
    \label{fig:all_data_dsi_trend}
\end{figure}

Blue curves in Fig.~\ref{fig:all_data_dsi_trend} represent time series of $A_f$ for different values $r$, the rate of change of airflow rate. Red dashed lines represent blowout time ($T_c$) obtained from experiments. We determine $T_c$ by measuring the global heat release rate. It is evident from those newly constructed time series that the value of $A_f$ remains very small initially and starts to increase as blowout is approached. The rise in the value of $A_f$ stops adjacent to blowout and experiences a rapid diminution after that. Such a behavior hints a singularity in $A_f$ accompanying a critical phenomenon, i.e., blowout. Note that the time at which maximum of $A_f$ occurs ($T_m$) and $T_c$ may not coincide. However, they remain within a single time window chosen while deriving $A_f$. In other words, $|T_c-T_m|<1~s$ because we are dealing with a time window of duration one second. Thus, for all practical purposes, we consider $T_c$ and $T_m$ are the same.

The system transitions to blowout earlier for a faster rate of change of airflow rates because the system reaches the critical equivalence ratio faster. However, the signature of increment in $A_f$ near blowout persists. The observed rise in $A_f$ preceding blowout itself is an interesting phenomenon having a significant prognostic value. A similar kind of behavior has been observed in the stock market index before market crashes~\cite{sornette2009_book}. The rest of the discussions are based on the analysis of data segments plotted in  Fig.~\ref{fig:all_data_dsi_trend}.

\subsection{Presence of oscillations preceding blowout in the time series data} \label{subsec:detection_of_oscillations}
At first, we examine whether the $A_f$ data preceding blowout possesses log-periodic oscillations intrinsically. Towards this purpose, we first perform a non-parametric test on $A_f$. The appearance of log-periodicity before blowout is determined following a method developed by Vandewalle et al.~\cite{Vandewalle_1999_EPJB} to detect the log-periodic component preceding a market crash. In their method, log-periodic patterns are confirmed by computing the upper ($y_{max}$) and lower ($y_{min}$) envelope functions of the market index ($y$). The upper envelope at any time $t$ is the maximum of $y$ until time $t$. Similarly, the lower envelope at $t$ is the minimum of $y$ from $t$ to the end of the time series. The two envelopes never meet for a simple oscillation such as sinusoidal oscillation, and their difference is constant. However, in the case of log-periodic power law (LPPL) oscillations, the two envelopes coincide at points where another period of the oscillation begins, and their difference becomes zero. Moreover, the appearance of such points increases as a system approaches a critical point following log-periodic oscillations. The two envelopes become equal because, at those points, they simultaneously attain a new value that they did not attain earlier. The difference between envelopes, also known as the running difference, comprises oscillations that correspond to log-periodic oscillations present in the time series. Vandewalle et al.~\cite{Vandewalle_1999_EPJB} found that oscillations obtained in this way accelerate as the critical time of the crash is approached and fitted the log-periodic term $\cos(\omega \log (t_c-t))$ to those oscillations. This method can highlight log-periodic oscillations.
\begin{figure}[ht]
    \centering
    \includegraphics[height=!,width=0.48\textwidth]{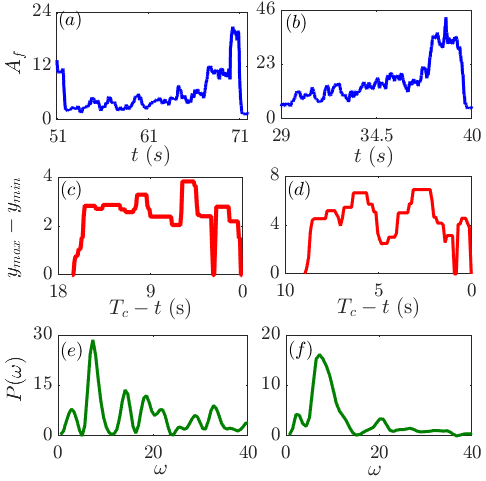}
    \caption{Characteristics of log-periodic oscillations for rates of change of airflow rates $0.4~g/s^2$ (left column) and $0.8~g/s^2$ (right column) as obtained from experiments. The time series of $A_f$ are shown in (a)-(b). Oscillations given by the difference between upper and lower envelope functions are shown in (c)-(d). Lomb periodogram of such oscillations are shown in (e)-(f). 
    }
    \label{fig:oscillations_characteristics}
\end{figure}

In Fig.~\ref{fig:oscillations_characteristics}, we show that the time series of $A_f$ (\ref{fig:oscillations_characteristics}(a), \ref{fig:oscillations_characteristics}(b)) for rate of change of airflow rates, $0.4~g/s^2$ (Fig.~\ref{fig:oscillations_characteristics} left column) and $0.8~g/s^2$ (Fig.~\ref{fig:oscillations_characteristics} right column), respectively. Oscillations as obtained from the running difference are shown in Fig.~\ref{fig:oscillations_characteristics}(c) and Fig.~\ref{fig:oscillations_characteristics}(d). In Fig.~\ref{fig:oscillations_characteristics}(a,b), for the time series of $A_f$ corresponding to the rate of change of airflow rates $0.4$ and $0.8 ~g/s^2$, oscillations are observed in between $51~s$ to $71~s$  and $29~s$ to $40~s$, respectively. The frequency of oscillations obtained from the running difference increases as $t \rightarrow T_c$, which indicates the existence of log-periodicity in the time series of $A_f$ prior to blowout. Note that here we treated the occurrence of the maximum of $A_f$ as $T_c$ to uncover the log-periodic oscillations. Next, we perform a spectral analysis of the computed running difference in $\log(T_c-t)$ to assess the indicated log-periodicity. Lomb power of the oscillatory components are shown in Fig.~\ref{fig:oscillations_characteristics}(e-f) whose high peaks confirm the existence of log-periodic oscillations in thermoacoustic systems prior to blowout.

The computation of the running difference serves as a sufficient condition for detecting log-periodic oscillations. Therefore, in order to strengthen our claim of the presence of log-periodic oscillations in turbulent thermoacoustic systems prior to blowout, we will perform a power law detrended measure of the time series of $A_f$. To serve the purpose, we need to remove an associated power law trend from the time series for analyzing the obtained residue. As a first step of the process, we will determine the power law associated with blowout. Note that the choice of a power law depends on how the observed variable evolves as the critical point ($T_c$) is approached. To understand the nature of power law and its exponents, we examine the growth rate of $A_f$ in the next subsection.

\subsection{Understanding the nature of the growth rate}

To understand the growth of a system, we generally plot the time series of a system variable $x$ in the $\log$-linear or semi-$\log$ scale ($t~vs.~\log x$). It is quite obvious that, the differentiation of $\log(x)$ with respect to time ($t$) i.e.,
\begin{equation}
    \frac{d}{d t}  \log(x) = \frac{1}{x} \frac{d x}{d t}
    \label{eq:deriv_log_p}
\end{equation}
provides the relative growth rate of $x$. We utilize semi-$\log$ plots to retrieve this relative growth rate of $x$ from slopes of the plotted time series without even knowing its explicit functional form. A constant slope indicates that the system grows exponentially. On the other hand, if the slope is increasing monotonically, the system is said to have a faster than exponential growth. A faster than exponential growth is accompanied by a singularity occurring at the critical time $t_c$~\cite{sornette_IJMPC:2002}. Such kind of a growth can be represented by the equation
\begin{equation}
    \frac{d x}{d t} = {x}^m,~m>1
    \label{eq:singularity_equation}
\end{equation}
yielding a solution of the form
\begin{equation}
    x(t)=x(0) ({1-\frac{t}{t_c}})^{-\frac{1}{m-1}}.
     \label{eq:solution_of_singularity_equation}
\end{equation}

Another way of quantifying the growth rate is to study the doubling time intervals $(\Delta t)_i$ over which the value of $x$ doubles~\cite{sammis_PNAS:2002}. Mathematically we can say, $x(t_{n}+(\Delta t)_n)=2 x_n$, where at the $n^{th}$ time $t_n$, $x_n=x(t_n)$. In case of an exponential growth of $x$, $(\Delta t)_i$ remain constant for all $i$. However, if $x$ doubles faster over a short time, the $n^{th}$ time interval $(\Delta t)_n$ can be written as,
\begin{equation}
    (\Delta t)_n=2^{-n(m-1)} (\Delta t)_0.
\end{equation}
Then, $(\Delta t)_n$ decreases following a geometric progression with the common factor $r=2^{-(m-1)}~(<1)$ and shrinks to zero as $t \rightarrow t_c$. Note that, computing the doubling of $x$ is not the only way to confirm a faster than exponential growth rate. In fact, this growth rate can be realized for any $\alpha$ times increment of $x$ over a time interval $(\Delta t)_i$ with $\alpha > 1$. In that case, $r$ can be generalized as
\begin{equation}
    r=\alpha^{-(m-1)},~\alpha>1,~m>1.
\end{equation}
In the present work, we consider $A_f$ as the system variable and set $\alpha=1.25~(1.35)$ for $r=0.4~g/s^2~(0.8~g/s^2)$ portraying a $25\%~(35\%)$ increment of $A_f$ to inspect $(\Delta t)_i$.
\begin{figure}[ht]
    \centering
    \includegraphics[height=!,width=0.5\textwidth]{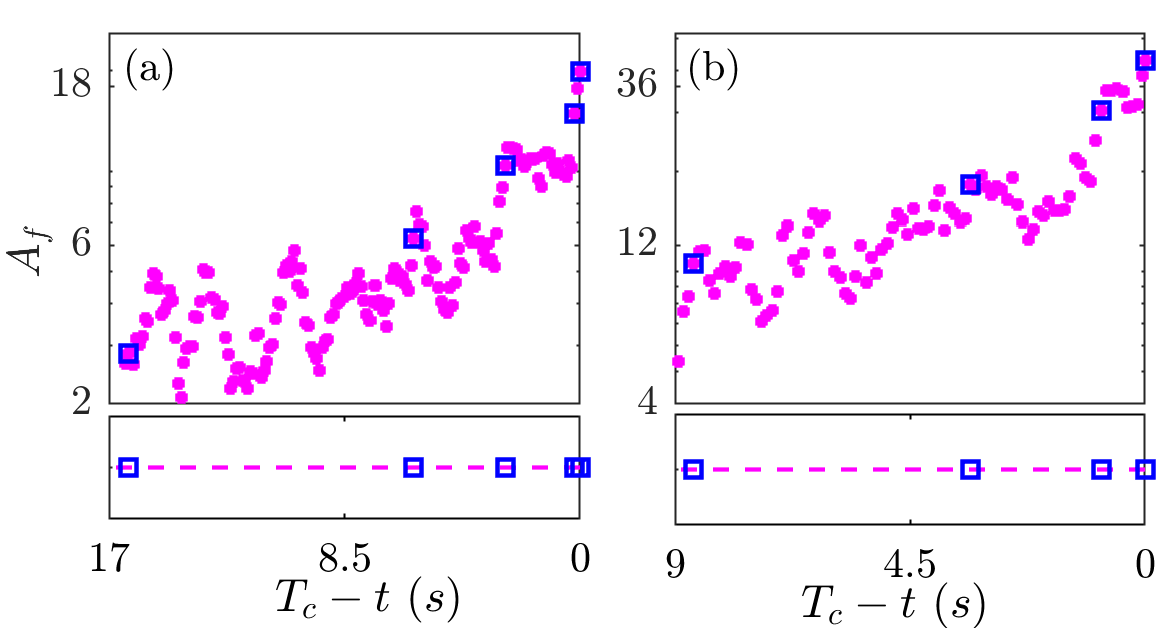}
     \caption{Semi-log plots of $A_f$ as a function of $T_c-t$ are shown for rates of change of airflow rates (a) $0.4$ and (b) $0.8~g/s^2$. Blue squares on $A_f$ curves represent successive $\alpha$ times increment of $A_f$ values. $T_c$ stands for the time at which the maximum of $A_f$ occurs. In the bottom row, the projection of blue squares on a $A_f=0$ line shows  the gradual diminution of time intervals which in turn, quantifies a faster than exponential growth of $A_f$ near blowout.}
    \label{fig:singularity_detection}
\end{figure}

 In Fig.~\ref{fig:singularity_detection}, we have shown the semi-$\log$ plot of the part of the time series of $A_f$ showing log-periodicity corresponding to different rates of change of airflow rates, $0.4~g/s^2$ in Fig.~\ref{fig:singularity_detection}(a), and (b) $0.8 ~g/s^2$ in Fig~\ref{fig:singularity_detection}(b). We estimate the growth function by calculating time intervals ($(\Delta t)_i$) between each $\alpha$-times increment in the $A_f$ value. Blue squares in Fig.~\ref{fig:singularity_detection} represent time instants forming $(\Delta t)_i$. We then project these time instances on the line corresponding to $A_f=0$ (shown in the bottom panel of Fig~\ref{fig:singularity_detection}) to unveil the fact that, for a fixed rate of change of airflow rate, time intervals ($(\Delta t)_i$) gradually decrease as the system approache blowout. Consequently, $A_f$ achieves a higher value in a short range of time close to blowout and becomes singular there. In each of these cases, it is apparent from the plot that the slope of $\log(A_f)$ is not a constant but changes as $t$ approaches the critical time to blowout $T_c$. Therefore, based on above discussions, we approximate $A_f$ as
\begin{eqnarray}
    A_f \approx (t_c-t)^{-\frac{1}{m-1}},
    \label{eq:hyper_exponential_power_law}
\end{eqnarray}
where $t_c$ denotes an approximated value of the experimentally obtained blowout time $T_c$. Earlier, Johansen et al.~\cite{johansen_2001} showed that the accelerated growth rate of an observable such as the world population, gross domestic product (GDP) of the world, and financial indices, can be approximated by a similar power law, leading to a superexponential behavior. Interestingly, the faster than exponential growth has also been realized in the spread of Covid-19 during the occurrence of the devastating Delta wave~\cite{covid_chaos_induja:2022}. 

Based on the approximation given by  Eq.~(\ref{eq:hyper_exponential_power_law}), we adopt a generalized power law representation for $A_f$ as,
\begin{equation}
    y(t)=A+B (t_c-t)^{-\frac{1}{m-1}}
    \label{eq:global_power_law_trend}
\end{equation}
 where, $A,~B$ are the linear and $t_c$ and $m$ are the nonlinear parameters. The variable $y$ and the parameter $t_c$ in Eq.~(\ref{eq:global_power_law_trend}) approximates $A_f$ and $T_c$, respectively. In the rest of the discussion, by $t_c$ we refer to the predicted time for blowout ($T_c$) obtained by curve fitting.  
 
\begin{figure}[t]
    \centering
    \includegraphics[height=!,width=0.48\textwidth]{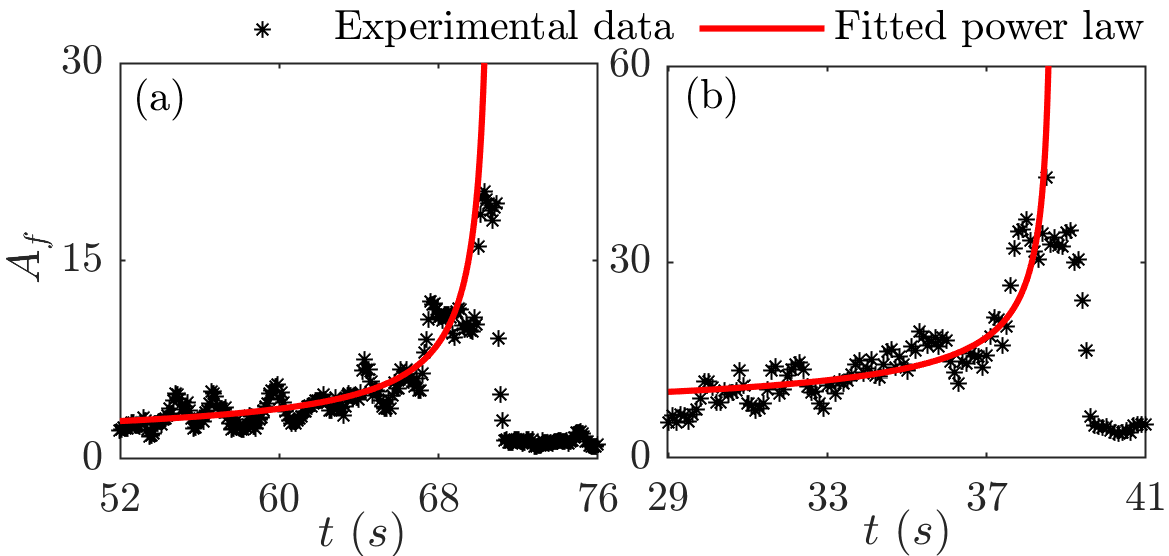}
    \caption{Power law fits to $A_f$ corresponding to rates of change of airflow rates (a) $0.4~g/s^2$, (b) $0.8~g/s^2$. Black markers represent $A_f$ obtained from the experiment. The curve in red represents Eq.~(\ref{eq:global_power_law_trend}) fitted to $A_f$. As $t \rightarrow t_c$ the fitted curve grows to infinity.
    The obtained parameter values are: $(a)~t_c=70.8,~m=2.28$ and $(b)~t_c=38.7,~m=2.86$.
    }
    \label{fig:fitted_law_1bym}
\end{figure}

 Next, we fit Eq.~(\ref{eq:global_power_law_trend}) to the time series of $A_f$ for different values of $r$. In Fig.~\ref{fig:fitted_law_1bym} we fit Eq.~(\ref{eq:global_power_law_trend}) to $A_f$ for rates of change airflow rates (a) $0.4~\text{and (b)}~0.8$ $g/s^2$ respectively. A power law fit gives $t_c=70.3~s$ and $38.7$ for $r=0.4~g/s^2$ and $0.8~g/s^2$, respectively. Since our purpose here is to identify the underlying log-periodicity, we choose the set of parameters such that $t_c$ remain very close to the actual time of occurrence of blowout. It is quite apparent from Fig.~\ref{fig:fitted_law_1bym} that $A_f$ values show oscillations with respect to the power law curve which could be log-periodic oscillations. In the next section we discuss about these oscillations in detail.
  \begin{figure}[ht]
    \centering
    \includegraphics[height=!,width=0.5\textwidth]{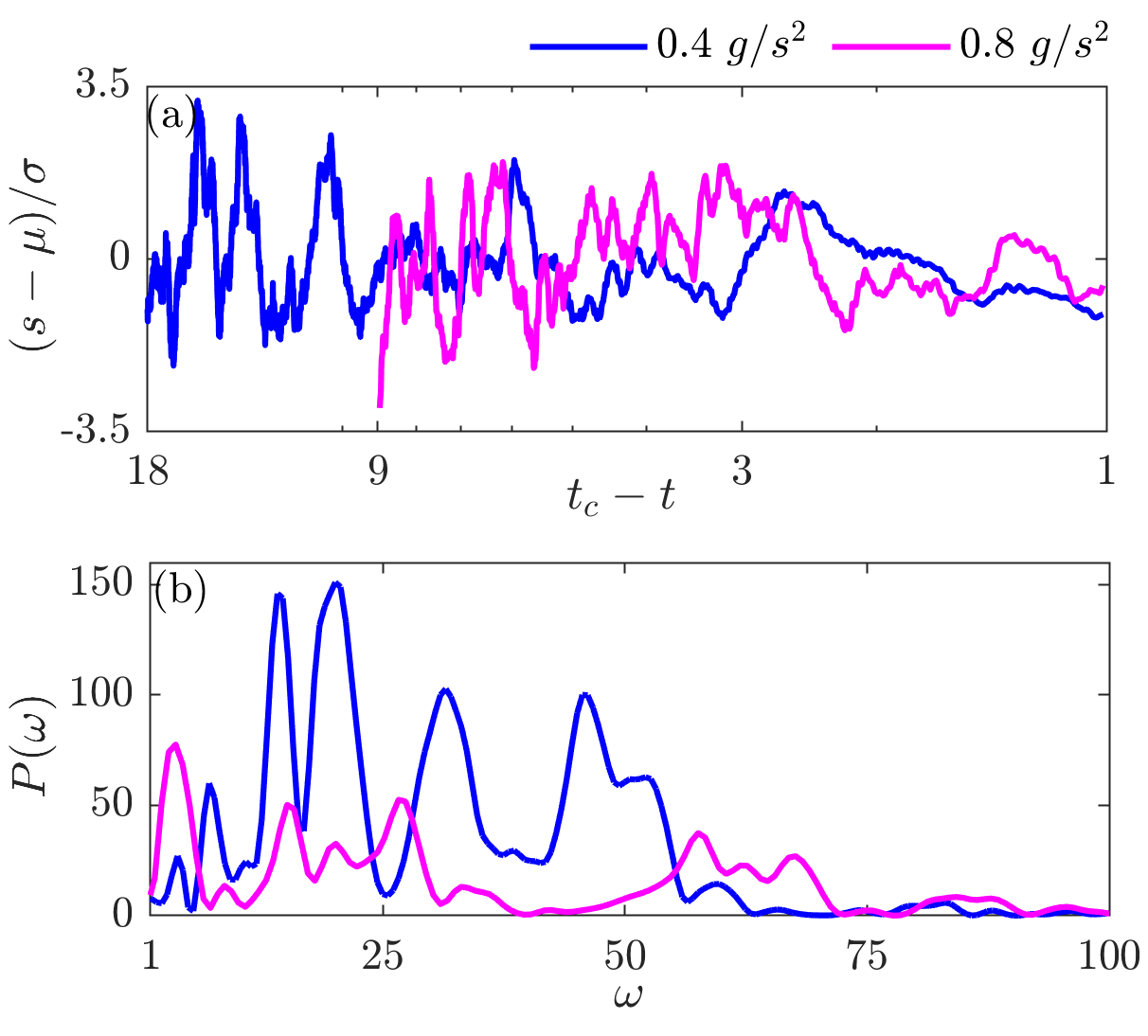}
    \caption{Power law detrended oscillations are shown in (a) for rates of change of airflow rates $0.4 ~g/s^2$ (red curve) and $0.8 ~g/s^2$ (blue curve) respectively. Normalized $s$ are plotted as a function of $\log (t_c-t)$ to get log-periodic oscillations. Lomb spectral power of log-periodic frequencies are shown in (b).  }
    \label{fig:detrended_law_1bym}
\end{figure}
 \subsection{Confirmation of log-periodic oscillations}
 A parametric detrended analysis given by Johansen et al.~\cite{Johansen_2000} can also detect log-periodic oscillations. We use this method to confirm qualitatively the presence of log-periodic oscillations already detected by the non-parametric method (discussed in section~\ref{subsec:log-periodicity}) in the time series of $A_f$. To serve that purpose, we first subtract parameter $A$ in equation (\ref{eq:global_power_law_trend}) from the actual $A_f$ data. Then, we detrend the power law $(t_c-t)^{-\frac{1}{m-1}}$ from this subtracted time series and perform Lomb periodogram analysis of the obtained residue ($s$) in $\log(t_c-t)$. The residue is given as:
 \begin{equation}
     s(t)=(A_f-A)/(t_c-t)^{-\frac{1}{m-1}},
 \end{equation}
 where $A,~m,~t_c$ are discussed in the preceding section. $\mu$ and $\sigma$ are the mean and the standard deviation of $s$. In Fig.~\ref{fig:detrended_law_1bym}(a) we have plotted the normalized $s$ with $\log (t_c-t)$ which shows some coarse oscillations.  PSD of the residue $s$ is shown in Fig.~\ref{fig:detrended_law_1bym}(b). $\omega= 2 \pi f$, is the log-angular frequency conjugate to $\log(t_c-t)$. 

From Fig.~\ref{fig:detrended_law_1bym}(b), we find that higher peaks in both cases occur when $\omega<50$. Hence, we interpret $\omega=50$ as upper limit for detecting fundamental log-periodic frequency. 

 Now, from the above discussions, it is evident that the preceding analysis confirms the presence of log-periodic oscillations prior to blowout in the $A_f$ data corresponding to different rates of change of airflow rates. These log-periodic oscillations confirm the presence of discrete scale invariance in thermoacoustic systems en route to blowout.
   \subsection{Log-periodic power law fitting} 
Equation~(\ref{eq:global_power_law_trend}) with log-periodic correction~\cite{Sornette_1996} is given as:
\begin{widetext}
\begin{equation}
y(t)=A+B (t_c-t)^{-\frac{1}{m-1}}+C (t_c-t)^{-\frac{1}{m-1}} \cos{(\omega \log(t_c-t)-\tau)},    \label{eq:global_log-periodic_function_old}
\end{equation}
\end{widetext}
where $\omega$ and $\tau$ are the angular frequency and the phase of the log-periodic oscillations. Equation ~(\ref{eq:global_log-periodic_function_old}) consists of seven parameters, out of which 3 are linear and 4 are nonlinear. The linear parameters are $A,~B,~C$, and the nonlinear parameters are $t_c,~m,~\omega,~\tau$. Due to the presence of a high number of nonlinear parameters, there are ambiguities in determining the parameters appropriately. Filiminov et al.~\cite{Filimonov_2013}, proposed a simplification of Eq.~(\ref{eq:global_log-periodic_function_old}) by incorporating the effect of $\tau$ into linear parameters and introduced two new parameters $C_1=C \cos(\tau),~C_2=C \sin(\tau)$ instead of $C$ and $\tau$.  In the present study, we follow the modified log-periodic equation proposed by Filiminov et al.~\cite{Filimonov_2013} given by
\begin{widetext}
\begin{equation}
    y(t)=A+B (t_c-t)^{-\frac{1}{m-1}}+ (t_c-t)^{-\frac{1}{m-1}} [C_1 \cos{(\omega \log(t_c-t))} + C_2 \sin{(\omega \log(t_c-t))}]. \label{eq:log-periodic_function}
\end{equation} 
\end{widetext}
This form of log-periodic Eq.~(\ref{eq:log-periodic_function}) has four linear parameters, namely, $A,~B,~C_1,~C_2$ and three nonlinear parameters $t_c,~m,~\omega$. Parameters are determined by minimizing the cost function
\begin{equation}
    z=\sum_{i=1}^{N} (p_i-y_i)^2 \label{eq:cost_function}
\end{equation}
over a time interval $[t_1,t_2]$ containing $N$ datapoints, where $p_i=A_f(t_i)$ is the value of data at $i^{th}$ time $t_i$, $y_i=y(t_i)$ is obtained from Eq.~(\ref{eq:log-periodic_function}). Among these seven parameters involved in the cost function, the linear and nonlinear parameters are solved separately. Initially, linear parameters, namely, $A,~B,~C_1,~C_2,$ are determined uniquely in terms of the nonlinear parameters by equating the first-order partial derivatives of the cost function $z$ with respect to linear parameters, to zero. Now $z$ becomes a function of nonlinear parameters only. Then $t_c,~m,~\omega$ are determined using nonlinear least square (NLS) method, the Nelder-Mead simplex method~\cite{Nelder_Mead_1965}. The characteristics of all seven parameters are discussed in Appendix~\ref{appendix1}.
\begin{figure*}[ht]
    \centering
    \includegraphics[height=!,width=\textwidth]{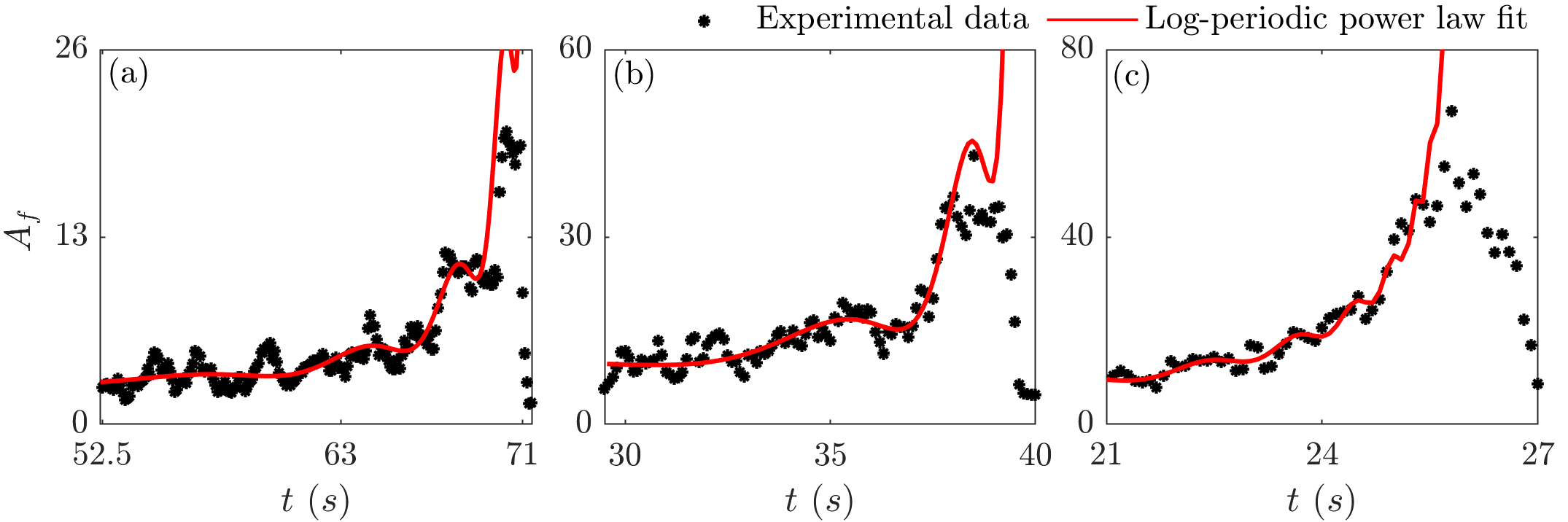}
    \caption{Log-periodic power law fit (solid red curves) to the experimentally obtained $A_f$ data (black markers) for different rates of change of airflow rates are shown ($0.4~g/s^2$ in (a), $0.8~g/s^2$ in (b) and $1.2~g/s^2$ in (c)). The nonlinear parameter values for these curves are: (a) $t_c=72.6,~m=1.6,~\omega=10.02$, (b) $t_c=40.4,~m=1.67,~\omega=7.67$ and (c) $t_c=25.9,~m=2.54,~\omega=14.5$}
    \label{fig:lppl_fit_to_data}
\end{figure*}

In Fig.~\ref{fig:lppl_fit_to_data}, we fit Eq.~(\ref{eq:log-periodic_function}) to the $A_f$ data for different rates of change of airflow rates ($r$), $0.4~g/s^2$ (in Fig.~\ref{fig:lppl_fit_to_data}(a)), $0.8~g/s^2$ (in Fig.~\ref{fig:lppl_fit_to_data}(b)) and $1.2~g/s^2$ (in Fig.~\ref{fig:lppl_fit_to_data}(c)). The fitted curves are shown in red while black markers represent data points obtained from the experiment. Predicted time for blowout or $t_c$ are 72.6 s (in Fig.~\ref{fig:lppl_fit_to_data}(a)), 40.4 s (in Fig.~\ref{fig:lppl_fit_to_data}(b)), and 25.9 s (in Fig.~\ref{fig:lppl_fit_to_data}(c)). The final point of the time interval $(t_2)$ selected for computation is $1~s$ earlier than $T_c$ in each case.
As the system approaches blowout, the fitted curve (shown in red) starts to grow and diverges. We also fit Eq.~(\ref{eq:log-periodic_function}) to several other data sets for different values of $r$. A comparison between mean of predicted blowout times ($t_c$) given by the pure power law (Eq.~(\ref{eq:hyper_exponential_power_law})), the LPPL (Eq.~(\ref{eq:log-periodic_function})), and the actual blowout time ($T_c$) for different values of the parameter $r$ are given in table~\ref{tab:LBO_comparison}. Here, mean of $t_c$ is the mean of 100 optimum $t_c$. The selection of these optimum values of $t_c$ are discussed in Appendix~\ref{appendix1} . We can clearly observe from the table~\ref{tab:LBO_comparison} that, predictions made by LPPL is better than the pure power law. Moreover, we calculate the percentage of relative error ($e=\frac{|t_c-T_c|}{T_c}$) for $t_c$ given by LPPL formulation. From table~\ref{tab:LBO_comparison}, we can find that the relative error percentage is less than $6\%$ for all the cases we have considered.

The wobbles in the LPPL curve in Fig.~\ref{fig:lppl_fit_to_data} are due to log-periodic oscillations. The local maxima or peaks of the curve are related to the underlying discrete scale-invariance property. Therefore, it is desirable that the predicted $A_f$ (or $y$) at those peaks are correlated to each other and may be associated with some hierarchical structure or collective behavior of microscopic components of the system which ceases to exist as blowout occurs. \\

Note that, the solution to the nonlinear parameters eventually gives the predicted time for critical phenomena ($t_c$), which corresponds to the time of blowout in the present case. Therefore, it will be interesting to know how early and how accurately can we predict blowout using Eq.~(\ref{eq:log-periodic_function}), which we will discuss in the next section.

\begin{widetext}
\begin{center}
\begin{table*}[h]
    \begin{tabular}{|c|c|c|c|c|c|c|c|}
    \hline
    $r~(g/s^2)$ & $t_c~(s)$ (power law) & $t_c~(s)$ (LPPL) & $T_c~(s)$ (LBO) &  $e~(\%)$ (LPPL)\\  \hline
  0.1 & 288.5 & 284.7 & 279.7 & 1.8 \\
  0.4 & 75.5 & 72.9 & 70.0 & 4.1 \\
  0.8 & 43.3 & 40.6 & 38.5 & 5.4 \\
  1.2 & 29.2 & 26.2 & 25.9 & 1.2 \\
  2.0 & 16.3 & 16.2 & 15.3 & 5.9\\
  \hline
    \end{tabular} 
    \caption{Experimentally determined and mean of predicted critical time for blowout for different rates of change of airflow rates. The predictions made by the LPPL is better than those by the power law.}
    \label{tab:LBO_comparison}
\end{table*}
\end{center}
\end{widetext}

\begin{figure*}[ht]
    \centering
    \includegraphics[height=!,width=0.9\textwidth]{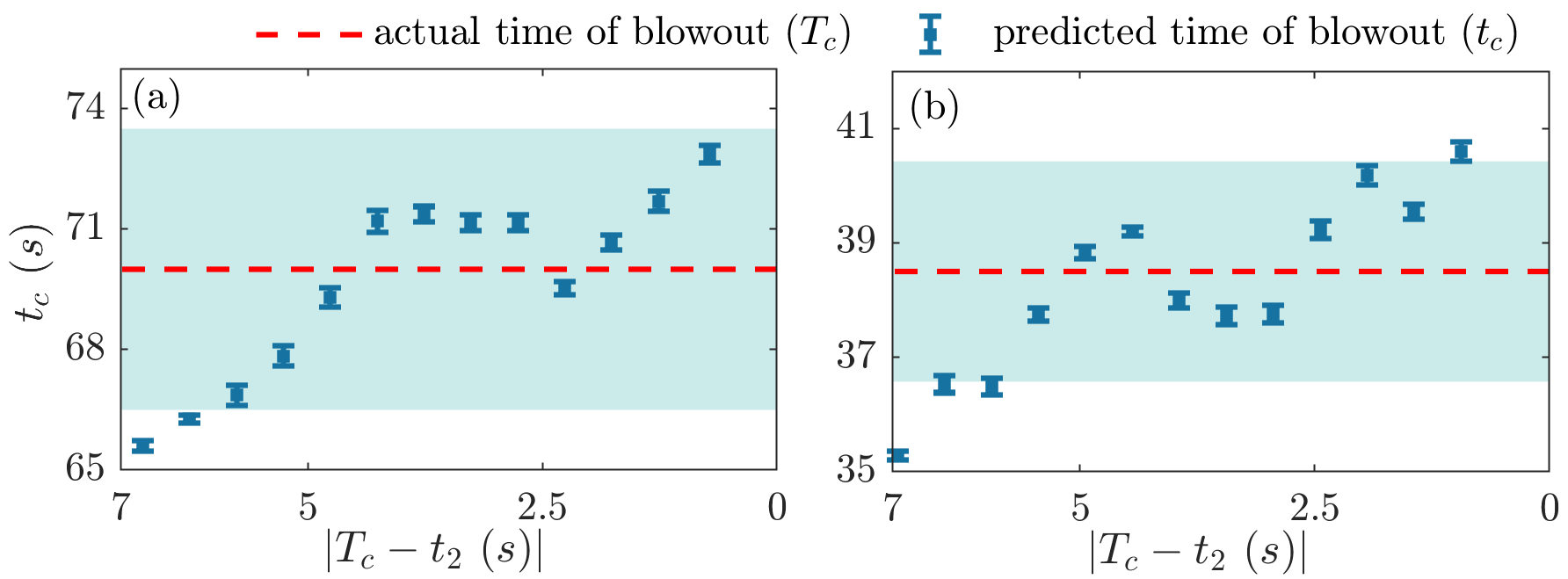}
    \caption{The predicted time to blowout ($t_c$) with error bars (blue markers) are plotted with respect to $T_c-t_2$ for different rates of change of airflow rates, $0.6~g/s^2$ in (a) and $0.8~g/s^2$ in (b). The red dashed line represents the experimentally obtained blowout time ($T_c$). The cyan-shaded region represents a threshold region of $T_c$.}
    \label{fig:t2_vs_tc_comparison_02}
\end{figure*}

\subsection{Early prediction of blowout}
Although $t_c$ is the predicted time of blowout, it is more reliable to consider the mean of $t_c$ for robust and accurate predictions instead of any particular instances of it. To serve the purpose, we first generate $5 \times 10^4$ realizations of $t_c,~m$ and $\omega$ for a sample space over the time window $[t_1,t_2]$. Then, we determine the $95\%$ confidence interval of $t_c$. We repeat the procedure for different sample spaces from a particular data set over $[t_1,t_2]$ by varying $t_2$. 
In Fig.~\ref{fig:t2_vs_tc_comparison_02}, we have plotted the predicted time of blowout ($t_c$) together with error bars (blue markers) as a function of $T_c-t_2$ for airflow rates $0.4$ and $0.8~g/s^2$. The absolute value of $T_c-t_2$ signifies how far is the actual onset of blowout from the present state of the system, i.e., the precedence of prediction. The cyan shaded region signifies a precision region of $T_c$ having a $5\%$ relaxation. If a $t_c$, or its error bar intersects the precision region, we can say that the associated interval ending at $t_2$ predicts $T_c$ well. In Fig.~\ref{fig:t2_vs_tc_comparison_02}(a), predictions together with its error remain within the precision region for $T_c-t_2 \leq 6~s$. In other words, better prediction can be given  up to 6 seconds earlier than the actual occurrence of blowout; after that, the predicted $t_c$ deviates from $T_c$. Similarly, in the case of rate of change of airflow rate $0.8~g/s^2$ (Fig.~\ref{fig:t2_vs_tc_comparison_02}(b)), we can approximate blowout up to $6.5~s$ earlier. Thus, the LPPL model, given by the Eq.~(\ref{eq:global_log-periodic_function_old}), can approximate the occurrence of blowout quite well. However, for fitting LPPL, we need enough data consisting at least one or two oscillations. Hence, the prediction could potentially fail for very fast rates of variation of control parameters. Since blowout occurs early with the increment of the rate of change of airflow rates, the distance between $t_1$, the starting point of the time interval over which fitting is performed, and $t_c$ decreases. Consequently, data points in the interval $[t_1,t_2]$ also decrease and become sparse for faster rates. When there is not enough data, predictions become challenging.

\section{Discussions and conclusions}
In this work, we characterize the dynamics of a thermoacoustic system close to lean blowout. We investigate the temporal variation of the maximum of coefficients of low-frequency spectrum with frequencies ($f$) from $5$ to $25~Hz$ denoted by $A_f$. $A_f$ is computed from pressure fluctuations ($p^\prime$) obtained from a laboratory-scale turbulent combustor for different rates of change of control parameter. We show that blowout can be explained, from the perspective of statistical physics, as a critical phenomenon. The value of $A_f$ after a specific time starts to increase with an oscillation of increasing frequency and continues until blowout. Thus, $A_f$ attains a finite-time singularity at the critical time where blowout occurs. We discover that such an oscillation in $A_f$ prior to blowout is the so-called log-periodic oscillation. The presence of log-periodic power law indicates an underlying discrete scale-invariance (DSI). Thus, we speculate that blowout can occur as a hierarchical phenomenon. Blowout has already been interpreted as the extinction of flames by the formation of global flame holes~\cite{nair_JPP:2005, nair_JPP:2007, chaudhuri_combustion_flame:2010}. The formation of local flame holes causes local flame extinction followed by reignition. The number and the frequency of extinction and reignition events increase as the system approaches blowout. Each extinction is a part of the precursory sequence of an even larger event with more local flame holes during such progression. At the critical time when blowout occurs, the collective behavior of local flame holes leads to the formation of a large flame hole that ultimately results in blowout. Therefore, the formation of such global flame holes can be considered a scaling-up process where each local extinction event is associated with intermediate scales in a geometric progression. However, further investigations are required to confirm the origin of log-periodicity close to blowout in thermoacoustic systems.

We also noticed that the growth of the amplitude of $A_f$ in the close vicinity of blowout follows a faster than exponential scaling law. Such a scaling law decorated with log-periodic oscillations gives a deterministic model to characterize $A_f$ prior to an impending blowout. Using the model, we are able to predict the occurrence of blowout well in advance for different rates of change of airflow rates. As per the authors' knowledge, no work has been done to predict blowout by identifying precursory log-periodic oscillations. The predicted blowout time obtained from the log-periodic model has an uncertainty of $5\%$ to the actual occurrence of blowout. Therefore, we expect that the model derived in the present work can serve the purpose of predicting impending blowout in lean operating combustors, enabling us to take control action in time to evade it.

\section*{Acknowledgement:} We express our sincere gratitude to Prof. W. Polifke and T. Komarek of TU Munich, Germany, for sharing the design of the combustor. We acknowledge the help from S. Thilagaraj, S. Anand and P. R. Midhun during the experiments. We thank K. Praveen, S. De, R. Rohit, S. Tandon, S. Srikanth, A. J. Varghese, R. Manikandan, A. Ghosh, K. V. Reeja for their valuable suggestions and comments on the draft. R.I.S. acknowledges the financial support from the Science and Engineering Research Board (SERB) of the Department of Science and Technology (grant no; CRG/2020/003051).
\appendix
\numberwithin{equation}{section}
\renewcommand{\theequation}{\thesection\arabic{equation}}
\setcounter{equation}{0}
\section{} 
\label{appendix1}

The log-periodic model given by Eq.~(\ref{eq:log-periodic_function}) is calibrated on the data within a time interval $[t_1,t_2]$. In the present analysis, we keep $t_1$ fixed and varied $t_2$. In order to determine the parameters, we start with a random initial guess of the nonlinear parameters $t_c,~m,~\omega$. At first, the linear parameters are determined uniquely by minimizing the cost function $z$ from Eq.~(\ref{eq:cost_function}) with respect to the linear parameters. In other words, equations
\begin{equation}
    \frac{\partial z}{\partial X}=0
\end{equation}.
 \begin{figure}[ht]
    \centering
    \includegraphics[width=0.9\textwidth]{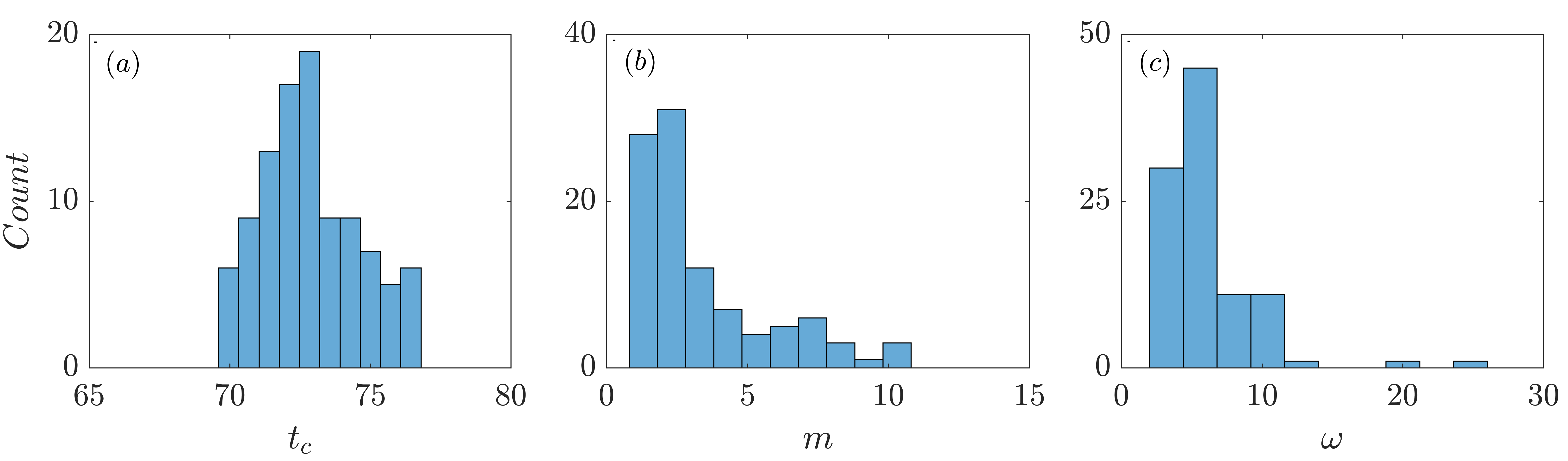}
    \caption{Histogram of nonlinear parameters: (a) $t_c$ , (b) $m$, and (c) $\omega$, computed over 100 realizations for rates of change of airflow rates $0.4~g/s^2$. Number of counts in each bin for each parameter are along vertical axis.}
    \label{fig:nonlinear_params}
\end{figure}
\begin{figure}[h!]
    \centering
    \includegraphics[height=!,width=0.6\textwidth]{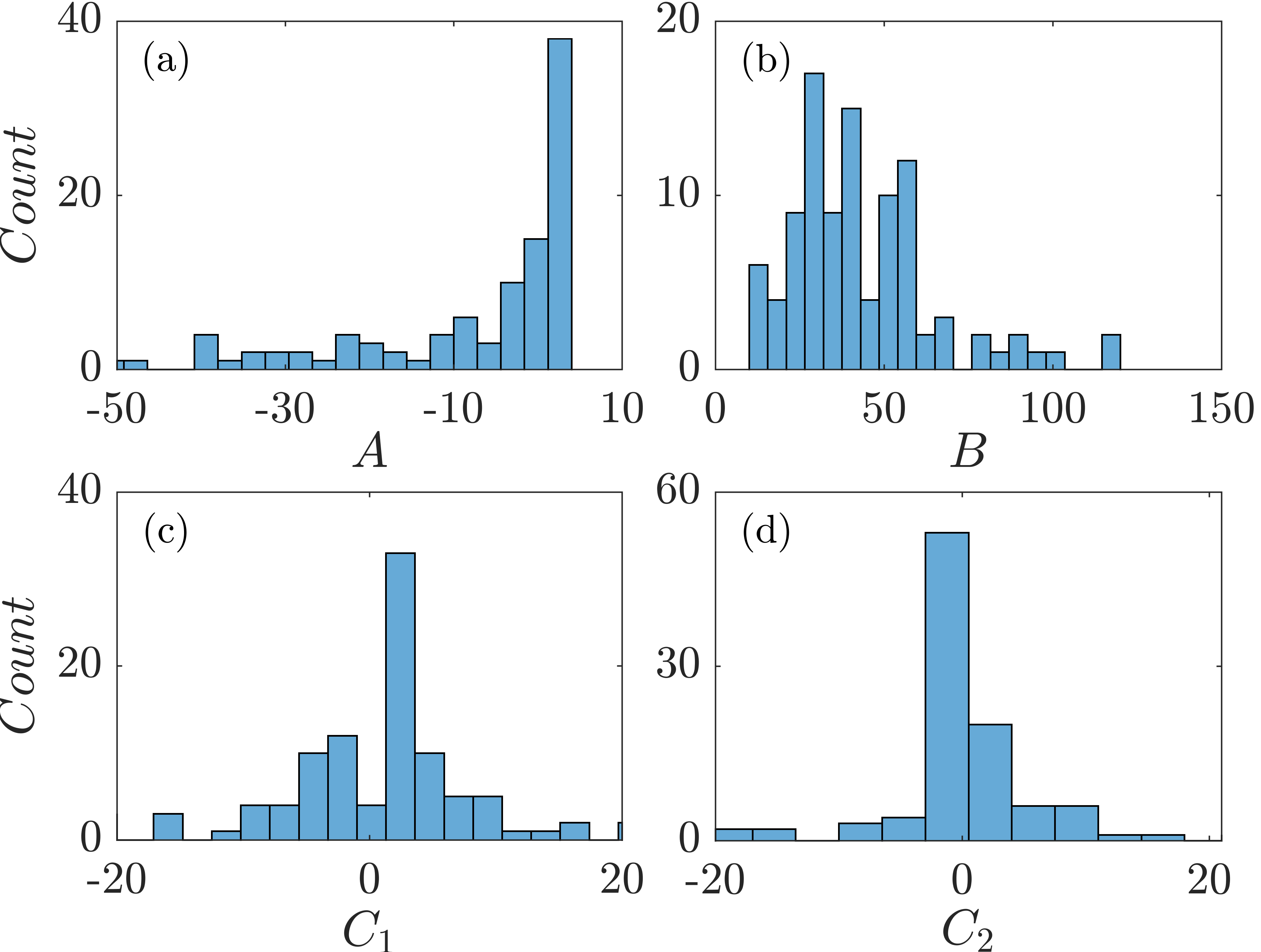}`   
    \caption{Histogram of linear parameters: (a) $A$, (b) $B$, (c) $C_1$, and (d) $C_2$, are shown for rates of change of airflow rates $0.4~g/s^2$. Ordinates represent number of counts in each bin for each parameter.}
    \label{fig:dist_linear_params}
\end{figure}
where, $X=A,~B,~C_1,~C_2$ are solved to get $A,~B,~C_1,~C_2$. Then we determine nonlinear parameters using nonlinear least square method. We use Matlab based package \textit{fminsearch} for determining $t_c,~m,~\omega$. We perform 100 realizations for each time interval examined in the present analysis. To increase the robustness of the process of determining parameters, each of the realizations consists of 500 iterations. We then compute sum squared error ($sse$) for all 500 iterations. The best set of parameters is considered to be associated with minimum $sse$ for each realization. Histograms of 100 such best set of parameter values are shown in in Fig.~\ref{fig:dist_linear_params} and Fig.~\ref{fig:nonlinear_params} for a rate of change of airflow rate of $0.4~g/s^2$. The histograms depict the fact that although the initial values of the nonlinear parameters are randomly selected, the computed final values are not scattered widely from their mean. 

\bibliography{log_periodicity}

\end{document}